\documentclass[sigconf, nonacm]{acmart}

\AtBeginDocument{%
  }

\usepackage{multirow}

\usepackage{algorithm}
\usepackage{algorithmic}
\usepackage{amsmath}

\begin{document}

\title{SF-RAG: Structure-Fidelity Retrieval-Augmented Generation for Academic Question Answering}
\author{Rui Yu}
\affiliation{%
  \institution{ Qilu University of Technology (Shandong Academy
of Sciences)}
  \city{Jinan}
  \state{Shandong}
  \country{China}
}

\author{Tianyi Wang}
\affiliation{%
  \institution{National University of Singapore}
  \city{Singapore}
  \country{Singapore}
}

\author{Ruixia Liu}
\affiliation{%
  \institution{ Qilu University of Technology (Shandong Academy
of Sciences)}
  \city{Jinan}
  \state{Shandong}
  \country{China}
}

\author{Yinglong Wang}
\affiliation{%
  \institution{ Qilu University of Technology (Shandong Academy
of Sciences)}
  \city{Jinan}
  \state{Shandong}
  \country{China}
}

\renewcommand{\shortauthors}{Rui Yu et al.}

\begin{abstract}
Efficient question-answering (QA) over extensive scientific literature is essential for evidence-based engineering decision-making. Retrieval-augmented generation (RAG) is increasingly applied to question-answering over long academic papers, where accurate evidence allocation under a fixed token budget is critical. However, existing approaches flatten papers into unstructured chunks, destroying the native hierarchical structure and forcing retrieval to operate in a disordered space. This produces fragmented contexts, misallocates tokens to non-evidential regions, and increases the reasoning burden for downstream language models.
To address these issues, we propose SF-RAG, an RAG framework that treats the native hierarchical structure of academic papers as a low-entropy retrieval prior. 
SF-RAG first inherits the native hierarchy to construct a structure-fidelity index, which prevents entropy increase at the source. 
It then designs a path-guided retrieval mechanism that aligns query semantics to relevant sections and selects high relevance root-to-leaf paths under a fixed token budget, yielding compact, coherent, and low-entropy retrieval contexts. 
In contrast to existing RAG approaches, SF-RAG avoids entropy increase caused by destructive preprocessing and provides a native low-entropy structural basis for subsequent retrieval.
We further introduce entropy-based structural diagnostics to quantify retrieval fragmentation and evidence allocation accuracy. 
Evaluations across three QA benchmarks show that SF-RAG significantly reduces retrieval fragmentation and improves evidence allocation. These structural benefits drive superior answer quality, establishing a scalable foundation for intelligent engineering document systems and future applications in technical specifications. \end{abstract}

\maketitle

\section{Introduction}
Academic literature serves as a critical knowledge source for engineering research and decision-making, with efficient question-answering over scientific publications being essential for evidence synthesis, technology assessment, and innovation planning in engineering practice. In these documents, an explicit hierarchy of title, sections, subsections, and paragraphs helps encode discourse intent and delineate evidence-bearing regions, providing a cognitive map for readers to localize, integrate, and verify claims~\cite{armbruster1987does,hebert2016effects}. To achieve automated reading and question answering over the increasingly growing amount of scientific publications and technical documents, retrieval-augmented generation (RAG) systems have been widely adopted. Recent studies in engineering informatics have demonstrated the potential of RAG in various domain-specific applications, such as building regulation interpretation~\cite{jiang2026building}, construction management~\cite{wu2025retrieval}, and smart manufacturing~\cite{wan2025empowering}. These systems typically segment document sources into flat chunks, followed by information retrieval via nearest-neighbor similarity in a global vector space~\cite{fan2024survey,gao2023retrieval,zhang2025survey,han2025rag,sudhi2024rag}. However, flat chunking weakens section boundaries, produces fragmented contexts, and elevates the risk of misallocating tokens away from evidence-bearing parts of the document under finite token budgets~\cite{yang2025superrag,xu2024unsupervised,ma2023query,cui2026llm}, which increases reasoning burden for downstream language models in practice~\cite{es2024ragas,salemi2024evaluating,lyu2025crud}.

\begin{figure}
    \centering
    \includegraphics[width=1.0\columnwidth]{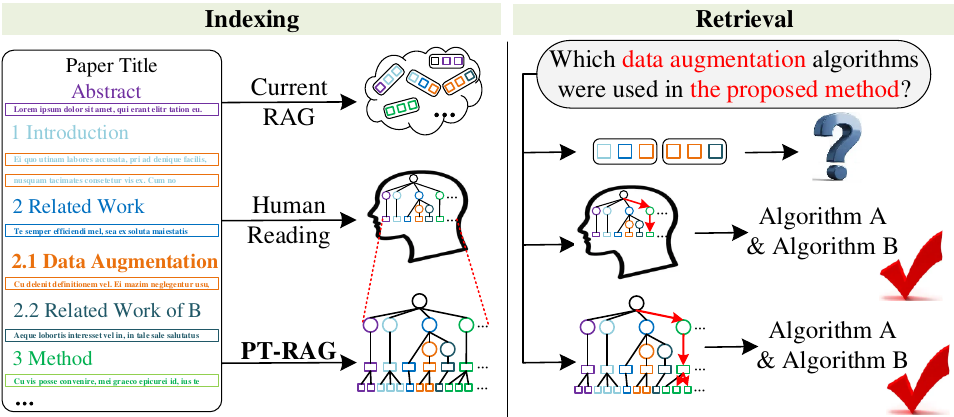}
    \vspace{-0.1in}
    \caption{Structure-Fidelity \& RAG. Current RAG methods flatten documents into unordered chunks, losing section signals and leading to fragmented, inaccurate retrieval. In contrast, humans naturally navigate papers via section hierarchies to localize evidence efficiently. Inspired by this, SF-RAG constructs a structure-fidelity index that preserves the native outline and performs path-guided retrieval, enabling precise, low-entropy context assembly under token budgets.}
    \label{toy_system}
    \vspace{-0.2in}
\end{figure}

Recent RAG approaches attempt to mitigate these issues by reconstructing semantic relation after flat chunking~\cite{zhang2025survey,han2025rag,yang2025superrag,xu2024unsupervised,cao2025neusym,jiang2024longrag,guo2024lightrag,edge2024local,sarthi2024raptor,fatehkia2024t,awotunde2026domain}. 
For instance, graph-based methods extract entities and relations to build concept or citation graphs~\cite{guo2024lightrag,zhang2024knowgpt,jimenez2024hipporag}, while hierarchical methods apply clustering and recursive summarization to form multi-level indices~\cite{sarthi2024raptor}.
Although these designs enhance retrieval accuracy, the reconstruction introduces noise and incurs additional computational overhead. More importantly, reconstructed structures are less effective than native hierarchical structures in supporting the location and interpretation of evidence by readers.
As a result, their ``destroy-and-rebuild'' paradigm struggles in a self-created high-entropy environment, which hinders reaching optimal solutions.

The law of entropy increase in information theory suggests that rising systemic uncertainty (information entropy) during information processing results in distortion and redundancy. As evidenced by prior methods, disrupting the native structure of documents markedly amplifies entropy, compelling retrieval to operate in a high-entropy chaotic environment. This leads to fragmented retrieval and evidence mismatch. Hence, effective knowledge retrieval should aim to suppress entropy increase at the source, rather than attempting to reconstruct semantic connections after structural disruption.
Accordingly, we propose a principle of ``structural-fidelity'': the design of RAG should preserve and leverage the native hierarchical organization of documents throughout the pipeline, thereby enabling accurate and efficient retrieval within a low-entropy and well-ordered semantic space.

Guided by this principle, We present SF-RAG, a structure-fidelity framework that explicitly treats the native hierarchical structure of academic papers as a low-entropy retrieval prior and assembles contexts by selecting coherent paths under a token budget. 
We develop two core mechanisms in SF-RAG, namely \textit{structure-fidelity indexing} and \textit{path-guided retrieval}. 
On the one hand, it constructs a structure-fidelity index by parsing the document and aligning content to the detected hierarchy, while it introduces structure-anchored summarization so that chunks created by splitting long sections retain clear positional and sequential relationships among them.
On the other hand, it enforces path-guided retrieval, where queries are first aligned to sections through semantic signals, followed by computing dual semantic relevance scores between the query and all chunks and summaries within the selected sections. Subsequently, contiguous root-to-leaf paths are prioritized based on token budget constraints, favoring those with consistently high and coherent relevance distributions to yield low-entropy, compact, and logically coherent retrieval contexts.
Unlike existing ``destroy-and-rebuild'' approaches, SF-RAG introduces a novel ``inherit-and-navigate'' paradigm.
Extensive experiments further demonstrate that SF-RAG mitigates evidence fragmentation and misalignment, assembles compact and coherent retrieval contexts, and thereby improves answer accuracy and quality. 
Main contributions of this work are as follows:
\begin{itemize}
\item Upon analyzing the underexplored role of native hierarchical structure in academic RAG, we propose SF-RAG, a structure-fidelity RAG that enables low-entropy and accurate evidence retrieval by preserving and leveraging the native document hierarchy.
\item To our knowledge, we are the first to introduce and implement the ``inherit-and-navigate'' paradigm in academic RAG. By inheriting the native structure of academic papers to build the retrieval index, SF-RAG avoids source entropy increase and reconstruction noise. Under strict token budget constraints, SF-RAG localizes information along coherent root-to-leaf paths, enabling efficient and precise retrieval within a low-entropy and well-ordered space. 
\item Experiments demonstrate that SF-RAG achieves superior performance by effectively reducing retrieval fragmentation and evidence misalignment, delivering more coherent and compact retrieval contexts, and thereby improving answer accuracy. This provides a methodological foundation for engineering document understanding systems.
\end{itemize}

\section{Related Work}
Efficient processing of scientific literature and technical documentation is a foundational capability for engineering information systems, supporting tasks ranging from regulatory compliance checking to technology monitoring. Recent research has explored various approaches to automate this process, evolving from early rule-based extraction to modern large language model (LLM) powered solutions. This section reviews existing methodologies in two key areas relevant to our work: academic paper comprehension systems and retrieval-augmented generation (RAG) techniques. We analyze their limitations in handling the hierarchical structure inherent in engineering knowledge sources and position our structure-fidelity approach within this landscape.
\subsection{Academic Paper Comprehension Systems}
Early systems for academic paper comprehension provide annotations, keyphrase extraction, citation highlighting, and heuristic organization~\cite{head2021augmenting,lee2016spotlights,kang2022threddy,kang2023synergi,august2023paper,chen2023marvista,fok2023scim,fok2023qlarify,kim2018facilitating,peng2022crebot}, preliminarily reduce barriers for readers and support navigation. However, later studies indicate that such systems do not fully capture discourse hierarchy or the dependencies that link claims to supporting evidence~\cite{armbruster1987does,hebert2016effects}. Therefore, recent systems integrate large language models for summarization~\cite{kang2023synergi}, question answering~\cite{peng2022crebot}, and outline generation~\cite{fok2023qlarify}. This integration increases generative capability and introduces risks such as hallucination and misalignment between answers and evidence in technical domains~\cite{fang2024enhancing,farahani2024deciphering,zhao2024longrag}. These observations motivate retrieval grounded generation that supplements model outputs with external evidence and that respects the structure of scientific writing.

\subsection{Retrieval Augmented Generation}
Retrieval-augmented generation (RAG) enhances large language model (LLM) outputs by grounding them with retrieved external information~\cite{ram2023context,fan2024survey}. The naive pipeline encodes a query into an embedding, segments documents into flat chunks, and retrieves top \textit{k} chunks by nearest neighbor similarity~\cite{gao2023precise,gao2023retrieval}. This design is straightforward, but it often ignores section boundaries and disrupts semantic continuity. As a result, context can be dispersed across many parts of a document and token budgets can be misallocated.

Subsequent methods introduce advanced, modular, graph structured, and hierarchical strategies~\cite{zhang2025survey,han2025rag,yang2025superrag,xu2024unsupervised,cao2025neusym,jiang2024longrag}. Advanced pipelines refine queries, perform multi-hop retrieval, add cross encoder re-ranking, and apply fusion techniques~\cite{gao2023precise,chan2024rq,yu2024rankrag}. Graph structured approaches extract entities and relations to build concept or citation graphs and then retrieve over graph neighborhoods~\cite{guo2024lightrag,edge2024local,jimenez2024hipporag}. Hierarchical approaches apply clustering and recursive summarization to build multi-level indices and tree structured views of content~\cite{sarthi2024raptor,fatehkia2024t}. These directions improve precision and interpretability for some tasks. However, they mostly retain the initial chunking step and reconstruct relationships after segmentation, a process that fundamentally adheres to the ``destroy-and-rebuild'' paradigm and thereby adds complexity and can introduce noise. Moreover, for the critical tasks of information localization, the reconstructed structure does not perform as well as the native structure in terms of both efficiency and accuracy.

In this work, we propose SF-RAG, an RAG method that introduces the ``inherit-and-navigate'' paradigm. It inherits document topology and performs path-guided retrieval under token constraints to enhance evidence alignment and reduce retrieval fragmentation and evidence misalignment.

\begin{figure*}[t]
  \centering
  \includegraphics[width=\linewidth]{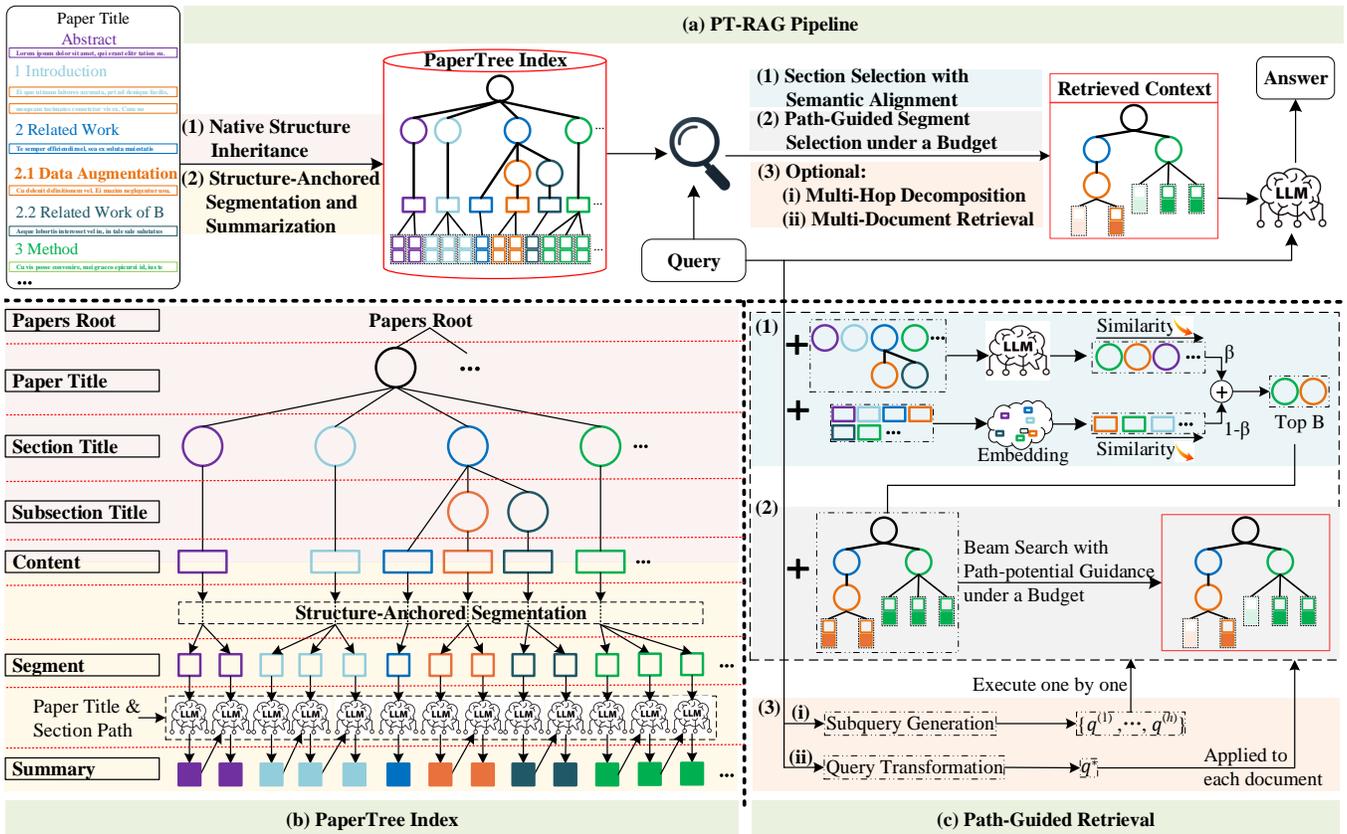}
  \caption{{Overview of the SF-RAG Framework. Our framework builds a structure-fidelity index via structure-anchored segmentation and summarization, enabling path-guided retrieval of coherent, low-entropy contexts for accurate answer generation. (a) SF-RAG Pipeline: illustrates the complete process from document parsing to answer generation. (b) Structure-fidelity Index: shows hierarchical organization of papers with contextualized summaries. (c) Path-Guided Retrieval: highlights query-adaptive path selection based on semantic relevance.}}
  \label{Method}
  \vspace{-0.15in}
\end{figure*}

\section{Method}

\subsection{Overview}
The SF-RAG pipeline is shown in Figure~\ref{Method}(a).
SF-RAG constructs a novel structure-fidelity index (Figure~\ref{Method}(b)) through native structure inheritance and structure-anchored segmentation and summarization. This process identifies the hierarchical organization of the input document, aligns content to section boundaries, and generates contextualized summaries that ensure segments produced by necessary segmentation of long sections retain well-defined positional and sequential relationships. This index provides a low-entropy basis for subsequent retrieval.
Then, the system introduces a path-guided retrieval mechanism (Figure~\ref{Method}(c)), which identifies relevant sections in response to a query using semantic signals.
Dual semantic relevance scores are then computed between the query and all chunks and summaries within the selected sections. Subsequently, contiguous root-to-leaf paths are prioritized based on token budget constraints, favoring those with consistently high and coherent relevance distributions, thereby yielding low-entropy, compact, and logically coherent retrieval contexts.
The retrieved context is fed into an LLM to generate accurate answers.

\subsection{Structure-fidelity Indexing}
\subsubsection{Native Structure Inheritance}
To support efficient and precise retrieval within a low-entropy space, we construct an index that directly inherits the native hierarchical structure of academic papers.  
The process begins by converting the input PDF into an intermediate Markdown representation that preserves line order and basic layout cues. All lines beginning with the hash symbol (\texttt{\#}) are initially treated as heading candidates, irrespective of their semantic role. This permissive strategy ensures high recall of genuine section headers at this stage.

The core of the construction is hierarchy inference, which jointly validates the authenticity of heading candidates and assigns consistent nesting levels. For each adjacent pair of heading candidates $(h_i, h_{i+1})$, a language model predicts their structural relationship among four possible classes: $h_{i+1}$ is a child of $h_i$, a sibling of $h_i$, a descendant of an earlier ancestor of $h_i$, or not a heading at all. The model outputs a probability distribution over these classes, and the label with the highest probability is selected as the local decision. A subsequent global reconciliation pass enforces tree validity by removing non-heading candidates and adjusting nesting levels to yield a well-formed hierarchical structure. When the model confidence for a local prediction falls below a predefined threshold, a rule-based fallback mechanism is activated. This fallback leverages conventional section names (e.g., ``Introduction'', ``Method'') and standard discourse patterns to assign levels in a robust manner. The final output is a rooted tree whose internal nodes correspond to the original section outline of the paper, and whose leaves are reserved for subsequent content segmentation.
This index, grounded in the native structure of academic documents, eliminates the entropy increase caused by flat chunking and avoids reconstruction-induced noise, thereby establishing a stable low-entropy foundation for downstream retrieval.

\subsubsection{structure-anchored Segmentation and Summarization}
Long sections are segmented necessarily into contiguous units to respect token budget constraints while preserving rhetorical boundaries. The segmentation algorithm groups consecutive sentences into segments of bounded length, respecting paragraph and list boundaries to avoid cutting across logical units. Each resulting segment inherits the full hierarchical path from the root to its parent section.

To minimize the entropy increase caused by segmentation, we generate a \textit{structure-anchored summary} $\sigma_i$ for segment $C_i$ using a prompt-conditioned language model. 
The summary is conditioned on four signals: the paper title $T$, the hierarchical path $H$, the current segment $C_i$, and the summary of the preceding segment $\sigma_{i-1}$ (or a special initialization token for the first segment of each section). This formulation ensures that each summary reflects not only local content but also its position in the discourse flow. Both the raw segment and its summary are stored as a leaf node in the structure-fidelity index, enabling dual-channel relevance scoring during retrieval.

\subsection{Path-Guided Retrieval}
To concentrate evidence along coherent paths and reduce fragmentation under a fixed token budget, retrieval proceeds in two stages, both guided by the structure-fidelity index.
First, the system identifies the most relevant sections by combining semantic alignment and embedding similarity. An LLM evaluates the correspondence between the query intent and the conventional role of each section title (e.g., whether the question pertains to experimental setup or theoretical analysis), yielding a semantic score \(F_{\mathrm{sec}}(q, s) \in [0,1]\). 
This fuses with a dense embedding similarity score \(E_{\mathrm{sec}}(q, s)\) between the query and the section body to produce a composite section relevance score, denoted as
\begin{equation}
R_{\mathrm{sec}}(q, s) = \alpha \, F_{\mathrm{sec}}(q, s) + (1 - \alpha) \, E_{\mathrm{sec}}(q, s),
\end{equation}
where \(E_{\mathrm{sec}}(q, s)\) computes as
\begin{equation}
E_{\mathrm{sec}}(q, s) = \cos(e(q), e(\mathrm{text}(s))),
\end{equation}
and \(e(\cdot)\) denotes the dense encoder. Here, \(\alpha \in [0,1]\) balances the semantic and embedding contributions. The top \(B\) sections retain as the retrieval scope to preserve recall.

Within the selected sections, evidence assembles via budget-constrained path selection. Each root-to-leaf path in the structure-fidelity index represents a coherent discourse thread. Segment-level relevance computes by fusing raw text and summary similarities,
\begin{align}
s_{\mathrm{seg}}(q, i) &= \cos(e(q), e(C_i)), \nonumber \\
s_{\mathrm{sum}}(q, i) &= \cos(e(q), e(\sigma_i)), \nonumber \\
S_i(q) &= \beta \, s_{\mathrm{seg}}(q, i) + (1 - \beta) \, s_{\mathrm{sum}}(q, i),
\end{align}
with cross-encoder re-ranking for precision. Here, \(\beta \in [0,1]\) weights the raw segment and summary similarities, \(C_i\) is the raw segment text, and \(\sigma_i\) is its structure-anchored summary. For every path \(p\), a relevance density computes as
\begin{equation}
U(p) = \sum_{i \in p} \frac{S_i(q)}{c_i},
\end{equation}
where \(c_i\) denotes the token cost of segment \(i\). The top \(P\) paths with highest density prioritize via beam search. Segments along these paths select in document order until the token budget \(T\) exhausts, guided by marginal relevance per token
\begin{equation}
\Delta(i) = \frac{S_i(q)}{c_i},
\end{equation}
under the optimization
\begin{equation}
\begin{aligned}
& \max_{A \subseteq \mathrm{Leaves}, \, \mathcal{P'} \subseteq \mathcal{P}} \sum_{i \in A} S_i(q) \\
& \text{s.t.} \quad \sum_{i \in A} c_i \leq T, \quad |\mathcal{P'}| \leq P, \quad A \subseteq \bigcup_{p \in \mathcal{P'}} p.
\end{aligned}
\end{equation}
where \(A\) is the set of selected segments, \(\mathcal{P'}\) is the set of selected paths, and Leaves denotes all leaf segments. This procedure inherently favors contiguous, section-consistent evidence and avoids cross-section jumps. The final context assembles by interleaving hierarchical paths, summaries, and raw segments in traversal order, preserving both logical flow and provenance.

\subsection{Optional Extensions}
\subsubsection{Optional Multi-Hop Decomposition}
For queries that require multi-hop reasoning, SF-RAG reuses the path-guided retrieval mechanism through query decomposition. A large language model rewrites the input query \(q\) into up to \(H\) subqueries \(\{q^{(1)}, \dots, q^{(h)}\}\), each representing a refined reasoning step. After retrieving evidence for \(q^{(t)}\), high-confidence entities extract from the context using threshold \(\tau_{\mathrm{ent}}\) and concatenate with \(q^{(t+1)}\) to guide subsequent retrieval while limiting error propagation. Decomposition stops if no new entities emerge for two consecutive steps or if \(H\) is reached. Each subquery applies the same section selection and path-guided retrieval under the global token budget. Evidence from all steps merges with deduplication and preserves original document order.

\subsubsection{Multi-Document Retrieval and Synthesis}
For multi-document queries, SF-RAG uses a query transformation mechanism to lower complexity while maintaining structural fidelity per document. A multi-document query such as ``Please analyze the research motivations presented in these papers and their interrelationships'' is first rewritten by a language model into a canonical single-document form: ``What is the research motivation of this paper?''. This transformed query is then applied independently to each target document by reusing the path-guided retrieval mechanism. The resulting evidence sets, each retaining its original hierarchical path and provenance, are aggregated and supplied together with the original query to the language model for final synthesis. This method avoids cross-document graph construction and ensures low-entropy retrieval within every document.

\subsection{Answer Generation}
Path-guided retrieval constructs the final context by preserving hierarchical paths, structure-anchored summaries, and raw segments from the original document. This structured representation, alongside the query, is delivered to the LLM to generate an answer, thereby improving the accuracy and quality of the answers.

\subsection{Probabilistic Structural Diagnostics}
\label{sec:prob_struct_diag}
To quantify how well retrieval aligns with the structural organization of academic papers, we introduce two entropy-based diagnostics derived from section-level probability distributions. Let \( S \) denote the set of sections in a document. For a given query, we construct a retrieved distribution \( \mathbf{r} \) by normalizing the fraction of tokens from each section included in the final context, weighted by segment relevance scores. Separately, we construct a ground-truth distribution \( \mathbf{g} \) by normalizing the number of annotated evidence tokens per section.

The first diagnostic, \textit{Section Entropy}, is defined as
\[
\mathrm{SE}(\mathbf{r}) = -\sum_{s \in S} \mathbf{r}(s) \log \mathbf{r}(s).
\]
Lower values indicate that the retrieval budget is concentrated in fewer sections, reflecting reduced fragmentation.

The second diagnostic, \textit{Evidence Alignment Cross Entropy}, is defined as
\[
\mathrm{EACE}(\mathbf{g} \,\|\, \mathbf{r}) = -\sum_{s \in S} \mathbf{g}(s) \log \mathbf{r}(s).
\]
Lower values indicate better allocation of tokens to evidence-bearing sections.

These metrics directly measure structural fidelity and provide a quantitative link between retrieval behavior and answer quality.

\section{Experiments}

\subsection{Experimental Setup}

\subsubsection{Datasets.} We conducted experiments on three scientific QA datasets: {QASPER} \cite{dasigi2021dataset}, {QASA} \cite{lee2023qasa}, and {M3SciQA} \cite{li2024m3sciqa}. These datasets are specifically chosen as they necessitate complex reasoning across multiple sections of academic documents.

\subsubsection{Baselines.}\footnote{To ensure fair comparison, baselines must return original text spans, and all models use the same reranking and generation modules with identical hyperparameter tuning.} We benchmarked SF-RAG against four representative baselines: the chunk-based {Naive RAG} \cite{gao2023retrieval}, the graph-based {GraphRAG} \cite{edge2024local} and {LightRAG} \cite{guo2024lightrag}, and the tree-structured\footnote{T-RAG is not included due to unavailable source code.} {RAPTOR} \cite{sarthi2024raptor}.

\subsubsection{Evaluation Metrics.} 
To evaluate model performance, we employ standard metrics for answer quality and evidence quality. All datasets provide human-annotated gold answers and evidence, enabling the use of: $\text{F}_1(\text{Answer})$~\cite{dasigi2021dataset}, BLEU~\cite{papineni2002bleu}, ROUGE~\cite{lin2004rouge}, and LLM-based selection (via \textsc{GPT-4o-mini})~\cite{edge2024local} for answer quality; $\text{F}_1(\text{Evidence})$~\cite{dasigi2021dataset} for evidence quality. Structural fidelity and fragmentation are measured by Section Entropy and Evidence Alignment Cross Entropy (Sec.~\ref{sec:prob_struct_diag}).

\subsubsection{Implementation Details}
We employ MinerU for PDF parsing \citep{wang2024mineruopensourcesolutionprecise} and use GPT-4o-mini as the backend for generation and semantic alignment \citep{achiam2023gpt}. Relevance scoring utilizes BAAI/bge-reranker-v2-m3 \citep{li2023making}. To ensure robust evaluation, we determined all hyperparameters via a small-scale grid search on the QASPER development set using Answer F$_1$ as the selection metric. The resulting configuration is kept constant across all datasets and experiments reported in this paper. Specifically, we set the section fusion weight $\alpha = 0.5$ and the segment balance weight $\beta = 0.8$. We limit fused segments to $M = 512$ tokens and consider up to $B = 2$ candidate sections during selection. Path-guided retrieval allows $P = 3$ paths under the token budget constraints. Optional multi-hop decomposition is configured for $H = 3$ reasoning hops. We observe that performance remains stable with neighboring values which indicates that SF-RAG is not sensitive to specific hyperparameter choices. All systems share identical answer generation templates and operate under matched retrieval and generation token budgets.

\begin{table*}[t]
\centering
\renewcommand{\arraystretch}{1.15}
\setlength\tabcolsep{6pt}
\begin{tabular}{c|l|c|cc|ccc|c|c|cc}
\hline
\hline
\multirow{3}{*}{\rotatebox[origin=c]{90}{Datasets}} & 
\multirow{3}{*}{Methods} & 
\multicolumn{7}{c|}{Answer} & 
\multicolumn{3}{c}{Evidence} \\
\cline{3-12}
& & \multirow{2}{*}{F$_1$} & \multicolumn{2}{c|}{BLUE} & \multicolumn{3}{c|}{ROUGE} & \multirow{2}{*}{LLM} & \multirow{2}{*}{F$_1$} & \multirow{2}{*}{SE} & \multirow{2}{*}{EACE} \\
\cline{4-8} 
& & & B-1 & B-2 & R-1 & R-2 & R-L & & & & \\ 
\hline

\multirow{7}{*}{\rotatebox[origin=c]{90}{QASPER}} 
& Naive RAG & 49.51\% & 42.30\% & 24.59\% & 50.92\% & 21.27\% & 47.25\% & 33.72\% & \textbf{54.28\%} & 2.27 & 2.28 \\
& GraphRAG & 33.47\% & 29.17\% & 11.17\% & 34.93\% & 5.35\% & 32.15\% & 20.68\% & 35.52\% & 1.45 & 1.51 \\
& LightRAG & 50.55\% & 44.51\% & 20.69\% & 52.44\% & 13.95\% & 47.98\% & 31.82\% & 25.62\% & 2.11 & 2.14 \\
& RAPTOR & 53.25\% & 45.52\% & 27.85\% & 55.09\% & 22.83\% & 47.88\% & 41.43\% & 40.09\% & \underline{1.29} & \underline{1.33} \\
& HiRAG & 50.16\% & 43.15\% & 23.40\% & 51.80\% & 19.50\% & 47.50\% & 32.50\% & 38.20\% & 1.92 & 2.08 \\
& HippoRAG2 & \underline{60.32\%} & \underline{52.45\%} & \underline{31.20\%} & \underline{61.50\%} & \underline{25.15\%} & \underline{56.80\%} & \underline{55.20\%} & 46.50\% & 1.31 & 1.47 \\
& SF-RAG\textsubscript{ours} & \textbf{67.09\%} & \textbf{60.12\%} & \textbf{34.87\%} & \textbf{67.95\%} & \textbf{27.56\%} & \textbf{64.84\%} & \textbf{69.83\%} & \underline{51.67\%} & \textbf{0.44} & \textbf{0.47} \\
\hline

\multirow{7}{*}{\rotatebox[origin=c]{90}{QASA}} 
& Naive RAG & 35.21\% & 28.68\% & 13.30\% & 39.35\% & 10.21\% & 31.72\% & 30.72\% & 12.63\% & 9.75 & 9.78 \\
& GraphRAG & 23.35\% & 17.68\% & 7.31\% & 26.51\% & 5.62\% & 20.68\% & 13.41\% & 12.91\% & 4.06 & 4.18 \\
& LightRAG & 37.01\% & 31.54\% & \underline{18.05\%} & 40.94\% & 7.78\% & 33.25\% & 32.13\% & 11.03\% & 4.75 & 4.90 \\
& RAPTOR & 37.50\% & 34.20\% & 16.37\% & 41.11\% & 8.35\% & 33.45\% & 34.48\% & \underline{13.85\%} & 3.78 & 3.90 \\
& HiRAG & 36.80\% & 31.10\% & 15.90\% & 40.50\% & 7.90\% & 32.80\% & 31.50\% & 11.50\% & 4.20 & 4.35 \\
& HippoRAG2 & \underline{39.55\%} & \underline{35.80\%} & 17.20\% & \underline{41.90\%} & \underline{10.50\%} & \underline{35.20\%} & \underline{38.50\%} & 13.20\% & \underline{2.50} & \underline{2.65} \\
& SF-RAG\textsubscript{ours} & \textbf{41.65\%} & \textbf{37.13\%} & \textbf{18.13\%} & \textbf{42.48\%} & \textbf{11.13\%} & \textbf{37.01\%} & \textbf{41.56\%} & \textbf{15.11\%} & \textbf{1.57} & \textbf{1.70} \\
\hline

\multirow{7}{*}{\rotatebox[origin=c]{90}{M3SciQA}} 
& Naive RAG & 29.03\% & 21.82\% & 14.38\% & 33.23\% & 18.60\% & 27.35\% & 26.94\% & 28.80\% & 4.28 & 4.29 \\
& GraphRAG & 16.27\% & 10.31\% & 6.92\% & 19.23\% & 10.11\% & 15.22\% & 5.13\% & 8.11\% & 6.46 & 6.65 \\
& LightRAG & 27.98\% & 21.04\% & 14.47\% & 31.97\% & 17.02\% & 26.02\% & 30.26\% & 4.53\% & 11.57 & 11.91 \\
& RAPTOR & 32.12\% & 24.95\% & 17.99\% & 33.34\% & 19.50\% & 28.64\% & 35.37\% & 28.33\% & 1.85 & 1.90 \\
& HiRAG & 30.50\% & 22.50\% & 15.80\% & 32.50\% & 18.10\% & 27.10\% & 31.20\% & 25.50\% & 5.10 & 5.25 \\
& HippoRAG2 & \underline{35.10\%} & \underline{27.50\%} & \underline{18.20\%} & \underline{37.50\%} & \underline{19.80\%} & \underline{32.50\%} & \underline{39.20\%} & \underline{32.10\%} & \underline{1.25} & \underline{1.30} \\
& SF-RAG\textsubscript{ours} & \textbf{37.68\%} & \textbf{30.04\%} & \textbf{18.71\%} & \textbf{41.29\%} & \textbf{20.12\%} & \textbf{36.26\%} & \textbf{42.57\%} & \textbf{35.81\%} & \textbf{0.66} & \textbf{0.72} \\
\hline
\hline
\end{tabular}
\caption{Main results on three benchmark datasets. SF-RAG consistently outperforms all baselines across all answer generation metrics. Best results are in \textbf{bold}, second best are \underline{underlined}.}
\label{tab:main_results}
\vspace{-0.2in}
\end{table*}

\subsection{Evaluation Results}
In this section, we present a thorough empirical evaluation of the proposed SF-RAG framework against leading baseline methods. We first evaluate the accuracy of the academic quality assurance benchmark in a question-answering setting, followed by an analysis of the quality and structural coherence of retrieved search results. Additionally, we perform ablation studies to analyze the contribution of key components, provide qualitative case studies, and discuss the broader impact of our findings.

\subsubsection{Answer Generation Quality}
We use F$_1$(Answer) as the key measure of answer quality under matched retrieval and generation budgets. As shown in Table~\ref{tab:main_results}, SF-RAG delivers consistent gains over strong baselines. On QASPER, F$_1$(Answer) increases from 60.32\% for the best baseline (HippoRAG2) to 67.09\% for SF-RAG, which is a gain of 6.77 points. On the other two datasets, SF-RAG maintains a consistent lead of approximately 2 to 3 points over the strongest competing methods. These improvements are obtained without increasing the budget or changing the reranker or generator, which indicates that the quality lift arises from the way SF-RAG assembles evidence for the generator rather than from additional capacity.

The pattern on QASPER is instructive. Naive RAG attains the highest F$_1$(Evidence) in Table~\ref{tab:main_results} (54.28\%) yet yields a much lower F$_1$(Answer) than SF-RAG. SF-RAG still achieves the best F$_1$(Answer) with fewer retrieved gold spans than Naive RAG on this dataset. This gap indicates that answer quality is limited not only by span overlap with gold but also by the structural coherence and density of the final context provided to the generator. Concentrating the token budget on a small number of section consistent paths produces contexts that are easier to integrate and cite, which improves the probability of generating the correct answer under the same budget.

We report two supporting observations with representative numbers. First, SF-RAG improves BLEU-1 on QASPER from 52.45\% for the best baseline to 60.12\%. Second, SF-RAG improves ROUGE-L on M3SciQA from 32.50\% for the best baseline to 36.26\%. These gains suggest that the generator benefits from more compact and well aligned contexts by producing answers that are both more accurate and linguistically closer to references. Together with the F$_1$(Answer) improvements, these results show that inheriting the native hierarchy and assembling coherent paths yields more usable evidence for the generator and translates into higher end performance across datasets.

\subsubsection{Retrieval Quality and Structural Diagnostics}
We evaluate retrieval quality with three complementary metrics in Table~\ref{tab:main_results}. F$_1$(Evidence) reflects the presence of gold evidence spans in the retrieved context. Section Entropy reflects how concentrated the context is across sections. Evidence Alignment Cross Entropy reflects how well the allocation concentrates on sections that contain gold evidence. Under fixed budgets, the joint outcome of these three metrics provides direct evidence of how SF-RAG assembles useful context for generation.

On QASPER, Naive RAG attains the highest F$_1$(Evidence) at 54.28\%, yet its Section Entropy and Evidence Alignment Cross Entropy are 2.27 and 2.28. In contrast, SF-RAG attains an F$_1$(Evidence) of 51.67\% with Section Entropy 0.44 and Evidence Alignment Cross Entropy 0.47. This joint result shows that SF-RAG retrieves a context that is both compact and correctly allocated to evidence bearing sections, while the baseline achieves higher span overlap through dispersed and misaligned retrieval. The improved answer quality of SF-RAG in Table~\ref{tab:main_results} is consistent with this interpretation.
On M3SciQA, SF-RAG improves all three metrics simultaneously. F$_1$(Evidence) increases from 32.10\% for the best baseline to 35.81\%. Section Entropy decreases from 1.25 to 0.66. Evidence Alignment Cross Entropy decreases from 1.30 to 0.72. This pattern indicates that SF-RAG concentrates the budget on fewer sections and allocates it to the sections that actually contain gold evidence, while also retrieving more of the annotated spans.

These representative cases illustrate how the three metrics together substantiate retrieval quality. Low Section Entropy without low Evidence Alignment Cross Entropy would indicate concentration on non evidence sections. High F$_1$(Evidence) together with high entropy diagnostics would indicate scattered matches that are hard to integrate. SF-RAG consistently achieves low Section Entropy and low Evidence Alignment Cross Entropy together with competitive or higher F$_1$(Evidence), which provides direct evidence that the retrieved contexts are compact, aligned with gold bearing regions, and more supportive of downstream generation under the same budget.

\begin{figure*}[t]
\setlength{\abovecaptionskip}{-0.1pt}
\centering
\includegraphics[width=\linewidth]{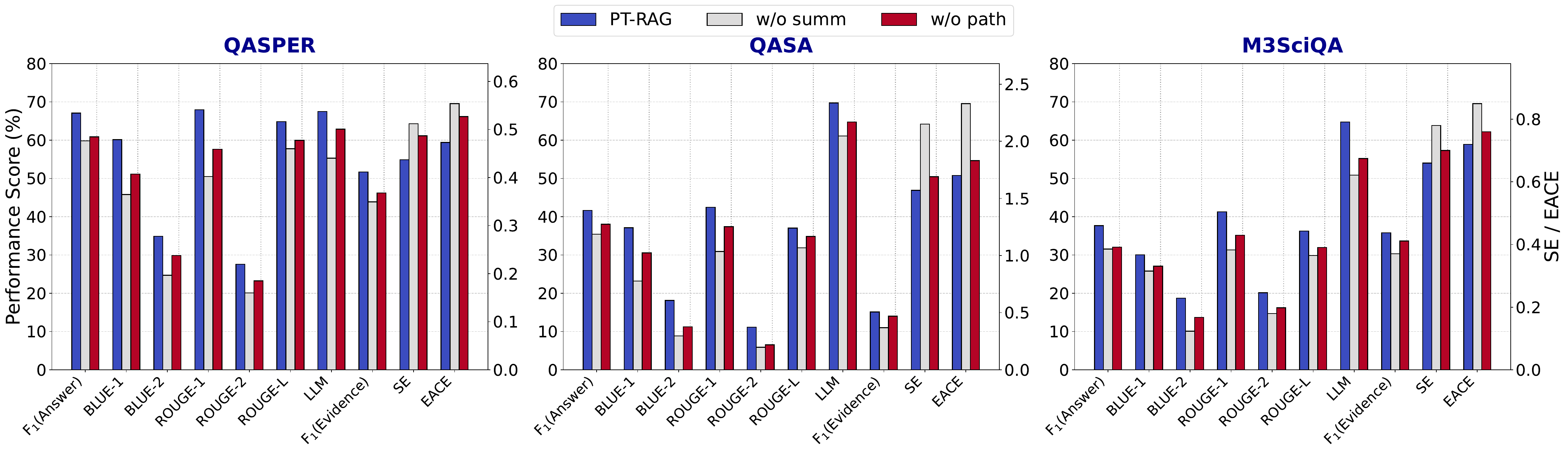}
\caption{Ablation study. Removing the contextual summary or the path-guided retrieval module reduces answer quality.}
\label{fig:Ablation_study}
\vspace{-0.1in}
\end{figure*}

\subsection{Ablation Study}
Figure~\ref{fig:Ablation_study} illustrates the individual contributions of the two core components in SF-RAG. The first ablation replaces the strongly coupled pair of the structure-fidelity index and the path guided retrieval with a global fragment retriever. These two parts are tightly integrated, and removing either one disables the other, therefore we report a single ablation for this pair, denoted as w/o path. The second ablation removes the structure anchored summaries while keeping the structure-fidelity index and the path guided retrieval unchanged, denoted as w/o summ. 

\noindent \textbf{w/o path.} Relative to SF-RAG, F$_1$(Evidence) consistently declines and both Section Entropy and Evidence Alignment Cross Entropy clearly increase across datasets. The rise in Section Entropy indicates that the budget becomes more dispersed over multiple sections rather than being concentrated along coherent routes. The rise in Evidence Alignment Cross Entropy indicates a larger allocation gap with respect to evidence bearing sections. These two shifts reflect the loss of structural guidance when retrieval no longer follows section consistent paths. F$_1$(Answer) drops accordingly in the same figure. In several settings the decrease in F$_1$(Answer) is more pronounced than the decrease in F$_1$(Evidence), which suggests that structural coherence and correct allocation amplify the utility of retrieved spans under a fixed budget.

\noindent \textbf{w/o summ.} Relative to SF-RAG, F$_1$(Evidence) declines and Evidence Alignment Cross Entropy increases markedly, while Section Entropy shows a smaller or more moderate increase than in the path ablation. This pattern indicates that, even when dispersion across sections is partially controlled by the path mechanism, removing structural anchoring weakens within section allocation and shifts tokens away from the most informative spans. F$_1$(Answer) decreases in parallel with these changes, which supports the interpretation that improved within section alignment is necessary for converting retrieved evidence into correct answers.

These two ablations reveal complementary roles. The coupled structure-fidelity index and path guidance primarily reduce dispersion across sections, which lowers Section Entropy and stabilizes answer quality. The structure anchored summaries primarily improve allocation within selected sections, which lowers Evidence Alignment Cross Entropy and supports answer generation. The full SF-RAG combines both effects, yielding the lowest entropy diagnostics together with the strongest F$_1$(Evidence), and this alignment is reflected by the highest F$_1$(Answer).

\begin{table*}[t]
\centering
\begin{tabular}{|l|p{15.5cm}|}
\hline
Query & What datasets were used for pre-training ViT, and what are the titles of their original papers? \\
\hline
Naive RAG &  The Vision Transformer (ViT) models ... The datasets utilized include:
\textbf{1. ImageNet}: This dataset ... The original paper reference is:
- Deng, J., ... (2009). \textbf{ImageNet: A Large-Scale Hierarchical Image Database.}
\textbf{2. ImageNet-21k}: This is ..., also-based on the original work by Deng et al. (2009).
\textbf{3. JFT}: This dataset ... The original paper reference for JFT is:
- Sun, C., ... (2017). \textbf{JFT: A Large-Scale Dataset for Big Visual Recognition.}
These datasets ...
\\
\hline
GraphRAG &  The Vision Transformer (ViT) models ...
The primary datasets include:
\textbf{1.JFT-300M}: This is a large-scale dataset ... [Data: Reports (40, +more)].
\textbf{2.ImageNet}: A widely recognized dataset for ... [Data: Reports (10, 21, 394)].
\textbf{3.CIFAR-10 and CIFAR-100}: These are standard datasets for ... [Data: Reports (26, 29)].
\textbf{Original Papers}
The original paper discussing the Vision Transformer models is titled:
- \textbf{\textit{An Image is Worth 16x16 Words: Transformers for Image Recognition at Scale}}.
This paper outlines ...
In summary, ...
\\
\hline
LightRAG & The Vision Transformer (ViT) model was primarily pre-trained using two significant datasets: \textbf{ImageNet-21k} and \textbf{JFT-300M}. Each of these datasets ...
\textbf{1. ImageNet-21k}
\textbf{Dataset Characteristics:}
- ImageNet-21k is a ...
\textbf{Related Paper:}
- The original work on ImageNet can be attributed to the paper titled \textbf{\textit{ImageNet: Large Scale Visual Recognition Challenge}}, which discusses ...
\textbf{2. JFT-300M}
\textbf{Dataset Characteristics:}
- JFT-300M is another ...
\textbf{Related Paper:}
- The JFT dataset is detailed in the paper titled \textbf{\textit{Scaling Up Visual and Vision-Language Pre-training with JFT}}, which captures ...
\textbf{Summary}
In summary, ...\\
\hline
RAPTOR & The Vision Transformer (ViT) models were pre-trained on the following datasets: \textbf{ImageNet}, \textbf{ImageNet-21k}, and \textbf{JFT-300M}. The titles of the original papers referenced in the context include:
\textbf{1. \textit{ViLBERT: Pretraining Task-Agnostic Visiolinguistic Representations for Vision-and-Language Tasks}} by Jiasen Lu et al., NeurIPS 2019.
2. The work by Dhruv Mahajan et al. on Big Transfer (BiT) models, although the specific title is not provided in the context.\\
\hline
HiRAG & The Vision Transformer (ViT) models were pre-trained on several datasets including \textbf{JFT-300M}, \textbf{ImageNet-21k}, and \textbf{ILSVRC-2012 ImageNet}. 
\textbf{1. JFT-300M}: ... consisting of over 300 million high-resolution images and 18,000 classes.
- \textbf{Original Paper}: \textit{``Revisiting Unreasonable Effectiveness of Data in Deep Learning Era''} by Sun et al., 2017
\textbf{2. ImageNet-21k}: ... containing 21,000 classes and over 14 million labeled images.
- \textbf{Original Paper}: \textit{``ImageNet: A Large-Scale Hierarchical Image Database''} by Deng et al., 2009
\textbf{3. ILSVRC-2012 ImageNet Dataset}: ... includes 1,000 classes and 1.3 million images.
- \textbf{Original Paper}: \textit{``ImageNet Classification with Deep Convolutional Neural Networks''} by Krizhevsky et al., 2012\\
\hline
HippoRAG2 & The datasets used for pre-training ViT are ImageNet, ImageNet-21k, and JFT300M. The titles of their original papers are: ImageNet (Deng et al., 2009), ImageNet-21k (Deng et al., 2009), and JFT (Sun et al., 2017).\\
\hline
SF-RAG & The datasets used for pre-training ViT are \textbf{ImageNet}, \textbf{ImageNet-21k}, and \textbf{JFT-300M}. The original paper for ImageNet and ImageNet-21k is \textbf{\textit{Imagenet: A large-scale hierarchical image database}} (Deng et al., 2009). The original paper for JFT-300M is \textbf{\textit{Revisiting unreasonable effectiveness of data in deep learning era}} (Sun et al., 2017).
\\
\hline
\end{tabular}
\caption{Qualitative comparison on a multi-hop query. SF-RAG is the only method that correctly retrieves and generates answer. }
\label{case_study}
\vspace{-0.15in}
\end{table*}

\subsection{Case Analysis and Human Evaluation}
We conduct a focused case study to examine complex reasoning patterns that are underrepresented in existing benchmarks. 
The study uses three representative papers\footnote{Paper 1: Attention Is All You Need. Paper 2: BERT Pre-training of Deep Bidirectional Transformers for Language Understanding. Paper 3: An Image is Worth 16x16 Words Transformers for Image Recognition at Scale} \cite{vaswani2017attention,devlin2019bert,dosovitskiy2020image} and seven query categories generated by GPT-4o-mini, comprising four single-document and three multi-document categories. Full case materials are provided in the project link. We conduct a blind human preference evaluation in which twenty annotators independently assess every query category, yielding 140 annotator–query evaluation instances in total. Each instance is judged on correctness, evidence sufficiency, and structural coherence. Across these 140 instances, SF-RAG more frequently returns answers that are judged correct with sufficient evidence and coherent structure under matched budgets. Table 2 presents one representative multi-hop query to illustrate typical reasoning patterns.

We present a representative multi-hop query in Table \ref{case_study}. Baselines show systematic weaknesses in handling precise citation mapping. Naive RAG retrieves the correct datasets but hallucinates the title for the JFT paper. GraphRAG introduces hallucinated citations unsupported by the text and misses the specific ImageNet-21k variant. LightRAG correctly identifies datasets but hallucinates a non-existent title for the JFT paper. RAPTOR identifies relevant datasets but fails to link them to the correct sources, citing unrelated work like ViLBERT. 
HiRAG correctly retrieves the JFT paper title, but for the ImageNet subset (ILSVRC-2012), it misattributes the source to the AlexNet model paper (Krizhevsky et al., 2012) rather than the original dataset paper. HippoRAG2 correctly identifies the authors and years for all datasets but fails to satisfy the user's specific request for paper \textit{titles}, providing only parenthetical citations. In contrast, SF-RAG correctly identifies ImageNet, ImageNet-21k, and JFT-300M, and accurately maps them to their respective original titles: \textit{``Imagenet: A large-scale hierarchical image database''} (Deng et al.) and \textit{``Revisiting unreasonable effectiveness of data in deep learning era''} (Sun et al.).
The result follows from path-guided retrieval that preserves the structural link between dataset entities and their bibliographic metadata. The retrieved context is compact and structurally coherent, which supports a grounded and complete answer that satisfies all constraints of the prompt.

\begin{table}[t]
\centering
\setlength\tabcolsep{6pt}
\begin{tabular}{l|cc|c}
\hline\hline
\multirow{2}{*}{Model} & \multicolumn{2}{c|}{Time (s)} & \multirow{2}{*}{Cost (USD)} \\
\cline{2-3}
& Indexing & Retrieval &  \\
\hline
Naive RAG & 0.00 & 12.39 & 0.01 \\
GraphRAG & 304.07 & 38.67 & 0.37 \\
LightRAG & 227.35 & 19.36 & 0.12 \\
RAPTOR & 413.02 & 26.91 & 0.03 \\
HiRAG & 731.67 & 21.18 & 0.26 \\
HippoRAG2 & 234.93 & 8.48 & 0.02 \\
SF-RAG$_{ours}$ & 30.98 & 17.74 & 0.02 \\
\hline\hline
\end{tabular}
\caption{Efficiency and cost under matched budgets.}
\label{tab:efficiency_cost}
\vspace{-0.2in}
\end{table}

\subsection{Efficiency and Cost Analysis}
\begin{figure}
\centering
\includegraphics[width=0.9\columnwidth]{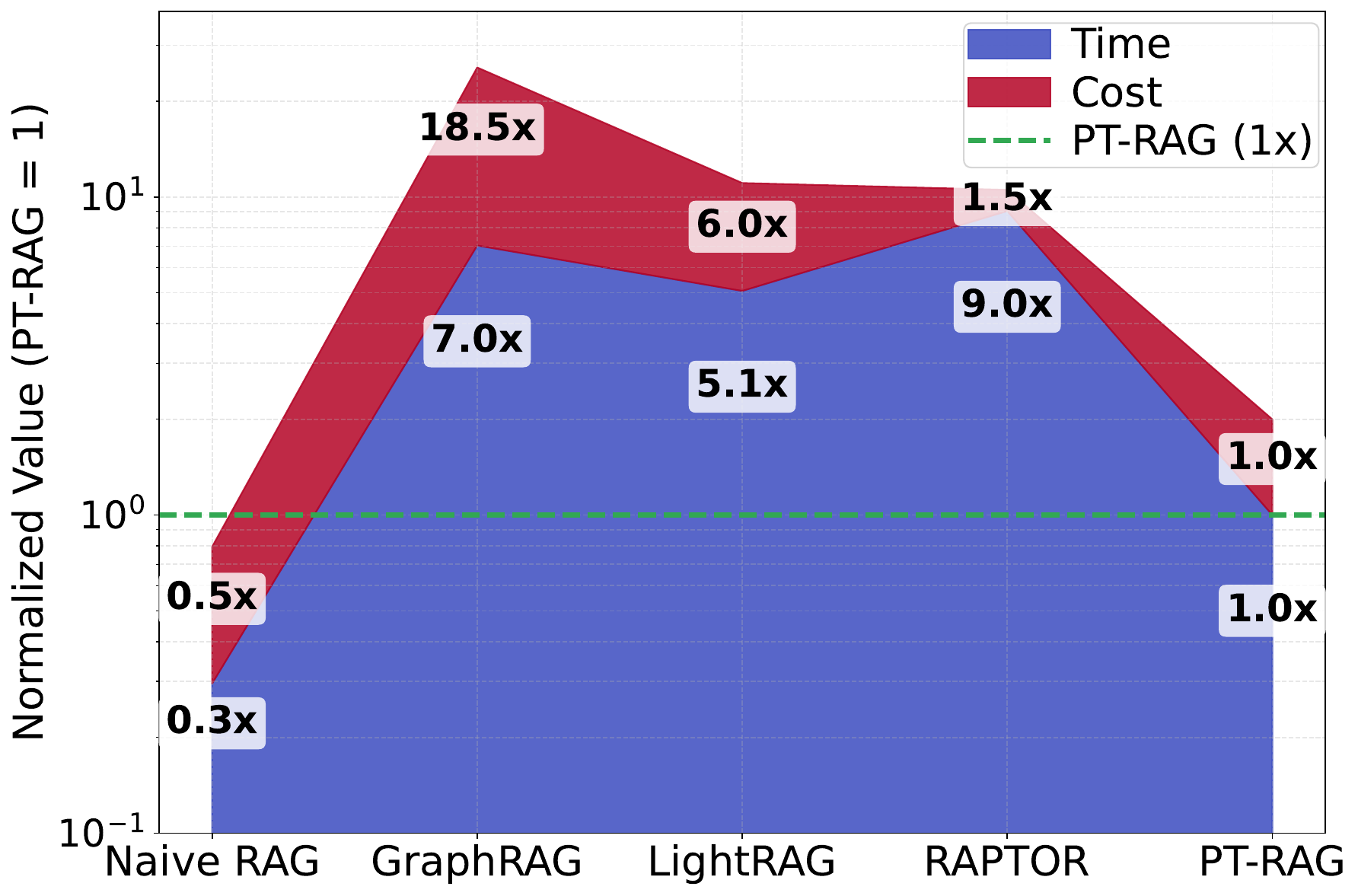}
\caption{Total efficiency and cost analysis under matched budgets. SF-RAG achieves lower latency and cost relative to advanced baselines.}
\label{fig:cost}
\vspace{-0.25in}
\end{figure}

We evaluate efficiency and cost under matched retrieval and generation budgets with identical hardware and concurrency settings. Figure~\ref{fig:cost} presents total time and monetary cost as ratios with SF-RAG as the baseline. Advanced baselines exhibit higher multiples in both total time and total cost. Table~\ref{tab:efficiency_cost} decomposes these totals into indexing and retrieval stages and shows the same trend.

The efficiency gains arise from two design choices. First, structure-fidelity indexing inherits the native hierarchy through outline parsing and per segment summarization. This workflow parallelizes and avoids global reconstruction such as graph construction and hierarchical clustering, which reduces indexing latency. Second, path guided retrieval narrows search to a small beam of relevant sections and a few coherent paths under the token budget. This restriction reduces candidate fragments, cross encoder reranking calls, and pre generation context length, which lowers retrieval latency and monetary cost.
Under the same budgets and with the same reranker and generator, SF-RAG operates at lower total time and lower total cost than advanced baselines. These reductions are achieved without loss of accuracy and are consistent with the compact, well aligned contexts reported in Table~\ref{tab:main_results}.

\subsection{Robustness and Generalizability of SF-RAG}

\begin{table*}[htbp]
\centering
\setlength\tabcolsep{2.8pt}
\begin{tabular}{l|l|ccccccc|c|cc}
\hline
\hline
\multirow{3}{*}{Model} & \multirow{1}{*}{LLM Size and} & \multicolumn{7}{c|}{Answer} & \multicolumn{3}{c}{Evidence} \\
\cline{3-12}
&Performance& \multirow{2}{*}{F$_1$} & \multicolumn{2}{|c|}{BLEU} & \multicolumn{3}{c|}{ROUGE} & \multirow{2}{*}{LLM} & \multirow{2}{*}{F$_1$} & \multirow{2}{*}{SE} & \multirow{2}{*}{EACE}\\
\cline{4-8}
&Difference& & \multicolumn{1}{|c}{BLEU-1} & BLEU-2 & \multicolumn{1}{|c}{ROUGE-1} & ROUGE-2 & \multicolumn{1}{c|}{ROUGE-L} &  & & \\
\hline
\multirow{3}{*}{{Naive RAG}} & GPT-4o-mini & 49.51\% & 42.30\% & {24.59\%} & 50.92\% & {21.27\%} & 47.25\% & {33.72\%} & 54.28\% & 2.27 & 2.28 \\
&QWEN2.5-7B& 41.61\% & 36.82\% & {16.87\%} & 43.71\% & {12.37\%} & 39.75\% & {41.17\%} & 56.72\% & 2.85 & 2.87 \\
&Difference& {↓}7.90\% & {↓}5.48\% & {{↓}7.72\%} & {↓}7.21\% & {{↓}8.90\%} & {↓}7.50\% & {{↑}7.45\%} & {↑}2.44\% & {↑}0.58 & {↑}0.59 \\
\hline
\multirow{3}{*}{GraphRAG}& GPT-4o-mini & 33.47\% & 29.17\% & 11.17\% & 34.93\% & 5.35\% & 32.15\% & 20.68\% & 35.52\% & 1.45 & 1.51 \\
&QWEN2.5-7B& 18.88\% & 12.83\% & {0.73\%} & 21.37\% & {0.80\%} & 15.74\% & {24.14\%} & 13.82\% & 3.78 & 3.95 \\
&Difference& {↓}14.59\% & {↓}16.34\% & {{↓}10.44\%} & {↓}13.56\% & {{↓}4.45\%} & {↓}16.41\% & {{↑}3.46\%} & {↓}21.70\% & {↑}2.33 & {↑}2.44 \\
\hline
\multirow{3}{*}{LightRAG}& GPT-4o-mini & {50.55\%} & {44.51\%} & 20.69\% & {52.44\%} & 13.95\% & {47.98\%} & 31.82\% & 25.62\% & 2.11 & 2.14 \\
&QWEN2.5-7B& 0 & 0 & {0} & 0 & {0} & 0 & {0} & 0 & N/A & N/A \\
&Difference& {{↓}{50.55\%}} & {{↓}{44.51\%}} & {{↓}{20.69\%}} & {{↓}{52.44\%}} & {{↓}{13.95\%}} & {{↓}{47.98\%}} & {{↓}{31.82\%}} & {↓}{25.62\%} & N/A & N/A \\
\hline 
\multirow{3}{*}{{RAPTOR}}& GPT-4o-mini & 53.25\% & 45.52\% & 27.85\% & 55.09\% & 22.83\% & 47.88\% & 41.43\% & {40.09\%} & 1.29 & 1.33 \\
&QWEN2.5-7B& 39.15\% & 33.03\% & {15.94\%} & 42.49\% & {12.17\%} & 35.72\% & {22.33\%} & 22.82\% & 2.18 & 2.26 \\
&Difference& {↓}14.10\% & {↓}12.49\% & {{↓}11.91\%} & {↓}12.60\% & {{↓}10.66\%} & {↓}12.16\% & {{↓}19.10\%} & {↓}17.27\% & {↑}0.89 & {↑}0.93 \\
\hline
\multirow{3}{*}{HippoRAG2}& GPT-4o-mini & 60.32\% & 52.45\% & 31.20\% & 61.50\% & 25.15\% & 56.80\% & 55.20\% & 46.50\% & 1.31 & 1.47 \\
&QWEN2.5-7B& 48.15\% & 40.50\% & {20.10\%} & 49.80\% & {16.50\%} & 45.20\% & {42.50\%} & 33.10\% & 2.05 & 2.18 \\
&Difference& {↓}12.17\% & {↓}11.95\% & {{↓}11.10\%} & {↓}11.70\% & {{↓}8.65\%} & {↓}11.60\% & {{↓}12.70\%} & {↓}13.40\% & {↑}0.74 & {↑}0.71 \\
\hline
\multirow{3}{*}{HiRAG}& GPT-4o-mini & 50.16\% & 43.15\% & 23.40\% & 51.80\% & 19.50\% & 47.50\% & 32.50\% & 38.20\% & 1.92 & 2.08 \\
&QWEN2.5-7B& 35.80\% & 29.50\% & {13.20\%} & 38.40\% & {9.80\%} & 33.10\% & {28.10\%} & 21.50\% & 3.15 & 3.35 \\
&Difference& {↓}14.36\% & {↓}13.65\% & {{↓}10.20\%} & {↓}13.40\% & {{↓}9.70\%} & {↓}14.40\% & {{↓}4.40\%} & {↓}16.70\% & {↑}1.23 & {↑}1.27 \\
\hline
\multirow{3}{*}{{SF-RAG}$_{ours}$}& GPT-4o-mini & 67.09\% & 60.12\% & 34.87\% & 67.95\% & 27.56\% & 64.84\% & 69.83\% & 51.67\% & 0.44 & 0.47 \\
&QWEN2.5-7B& 56.16\% & 47.45\% & {23.39\%} & 57.03\% & {17.80\%} & 55.47\% & {73.66\%} & 41.62\% & 0.89 & 0.95 \\
&Difference& {↓}10.93\% & {↓}12.67\% & {{↓}11.48\%} & {↓}10.92\% & {{↓}9.76\%} & {↓}9.37\% & {{↑}3.83\%} & {↓}10.05\% & {↑}0.45 & {↑}0.48 \\
\hline
\hline
\end{tabular}
\caption{Performance comparison on answer generation and evidence retrieval under GPT-4o-mini and Qwen2.5-7B. SF-RAG achieves the strongest performance and shows smaller degradation under capacity reduction. LightRAG outputs are invalid under Qwen2.5-7B because graph construction relies on structured triples that were not reliably produced by the weaker model.}
\label{tab:Compatibility}
\vspace{-0.2in}
\end{table*}

\begin{table*}[t]
    \centering
    \begin{tabular}{p{2cm}p{6cm}p{8cm}}
        \toprule
        \toprule
        \textbf{LLM Operation} & \textbf{Description} & \textbf{Fallback Mechanism} \\
        \midrule
        Section Title Reordering & Leverages LLM to reorder chapter titles semantically and logically, optimizing index structure. & If LLM fails to reorder effectively, the system automatically defaults to treating the paper title as primary and all other titles as secondary, ensuring structural integrity. \\
        \midrule
        Segment Summary Generation & Generates summaries for small segments with limited content, focused themes, and rich contextual information. & \textbf{No specific fallback mechanism}, as the task is well-defined and content-limited, resulting in low reliance on LLM capabilities and minimal risk of failure. \\
        \midrule
        Identification of Relevant Titles & Utilizes LLM to identify chapter titles highly relevant to the user query, quickly narrowing the retrieval scope. & If LLM identification fails, the system switches to an \textbf{embedding-vector-based semantic similarity strategy}, calculating similarity between the query and all titles. \\
        \midrule
        Query Decomposition & Decomposes complex or multi-intent queries into finer-grained sub-queries to enhance retrieval precision. & If decomposition fails (i.e., multiple sub-queries cannot be extracted from LLM output), the \textbf{LLM's complete output is directly used for retrieval}, ensuring an uninterrupted process. \\
        \midrule
        Entity Extraction & Uses LLM to extract entities from the target text that are relevant to the next query statement. & If extraction fails, the system does not update the next query statement and proceeds \textbf{with the original next query statement for subsequent retrieval}. \\
        \midrule
        Query Transformation & Transforms Multi-document query into Single-document query for retrieval mechanism reuse. & \textbf{No fallback mechanism}, as the LLM's complete output directly serves as the transformed query, meaning no failure in transformation. \\
        \bottomrule
        \bottomrule
    \end{tabular}
\caption{LLM operations in SF-RAG and fallback mechanisms that preserve progress under imperfect outputs.}
\label{tab:LLM_Operations}
\end{table*}

This evaluates robustness and generalizability under three axes that correspond to the design of SF-RAG and to the failure modes analyzed in the main text. The axes are structural priors in academic documents, reliability of the PDF parser, and sensitivity to LLM capacity. The goal is to test whether SF-RAG maintains low fragmentation, accurate evidence allocation, and low residual alignment error when upstream conditions vary.

\noindent \textbf{Use of Structural Priors.}
SF-RAG is built to inherit the conventional hierarchy in academic writing and to operate effectively even when only a coarse outline is available. In practice a single level of section headers is sufficient to construct structure-fidelity index and to enable path-guided retrieval. This choice reduces preprocessing complexity and avoids destructive reconstruction. It also mitigates dispersion across sections because retrieval is localized within the author defined structure. The resulting context is more compact and easier to align with answer bearing regions under a fixed token budget.

\noindent \textbf{Reliability of the PDF Parser.}
structure-fidelity index construction begins with structured content extraction from PDF. We use MinerU, an open source framework designed for academic documents. On the three datasets used in this study, which include 631 papers, all PDFs were converted to structured Markdown without observed conversion failures. MinerU identified 11,413 section headers in total, which corresponds to an average of 18.08 headers per document. Two edge cases illustrate typical behavior. On a 61 page paper with complex layout, MinerU identified 76 headers while manual inspection found 72. The four errors were a table title and three prompt examples. These were corrected during header validation. On a 2 page paper with simple layout, MinerU correctly identified all 7 headers. These observations indicate that MinerU outputs are sufficiently accurate and stable for the requirements of SF-RAG, and that the validation stage corrects most residual errors.

\noindent \textbf{Robustness to LLM Capacity Reduction}
Many RAG pipelines rely on the LLM to infer latent structure and to reconcile dispersed fragments, which increases sensitivity to model capacity. SF-RAG decouples these burdens by using structure-fidelity index and structure-anchored summaries to provide organized and position aware context. Table \ref{tab:Compatibility} reports results when GPT-4o-mini is replaced with Qwen2.5-7B. SF-RAG retains higher answer quality and evidence quality relative to baselines and exhibits a smaller performance drop across most metrics. Naive RAG and RAPTOR show larger degradation in answer metrics. GraphRAG shows large drops due to dependence on entity and relation extraction. LightRAG fails to construct a usable graph under the smaller model, which prevents downstream retrieval.

The architecture of SF-RAG contributes to this stability in three ways. First, structure-fidelity index and path-guided retrieval supply contiguous and section consistent evidence, which reduces fragmentation and eases integration. Second, LLM tasks are well scoped and local, for example header validation and segment summarization, which lowers variance across model sizes. Third, fallback mechanisms replace uncertain decisions with deterministic alternatives, for example section ranking by embeddings when semantic alignment is unreliable. Together these properties reduce misallocation of tokens to non bearing sections and lower residual alignment error under capacity reduction.

\subsection{Generalization Analysis}
To evaluate the generalization capability of SF-RAG beyond academic literature, we conducted additional experiments on the DOCBENCH dataset \cite{zou2025docbench}. This benchmark comprises five distinct document domains: Academia, Finance, Government, Laws, and News, representing a diverse spectrum of structural complexity from highly standardized legal codes to loosely organized news articles. Unlike our primary datasets, DOCBENCH lacks fine-grained evidence annotations required for calculating entropy metrics (SE and EACE). Therefore, we focus solely on end-to-end answer generation performance (F$_1$). The primary objective of this experiment is to verify whether SF-RAG, originally designed with hierarchical priors for academic papers, maintains robustness when applied to real-world documents characterized by non-standard structures, irregular formatting, and varying levels of hierarchical noise.

As shown in Table~\ref{tab:docbench_f1}, SF-RAG consistently outperforms all baselines across all five domains, achieving an overall F$_1$ score of 62.12\% and surpassing the strongest baseline HippoRAG2 by an average of 5.1\%. Notably, performance gains are positively correlated with document structural integrity. The most significant improvements are observed in highly structured domains (Academia: +7.4\%, Laws: +6.6\%), confirming that our probabilistic tree construction effectively exploits explicit hierarchical signals. Crucially, even in domains with weak or noisy structures (Finance: +2.9\%, News: +2.2\%), SF-RAG maintains a stable advantage over baselines. This demonstrates that while our method benefits most from clear hierarchy, its soft clustering and confidence-weighted retrieval mechanisms prevent performance collapse in unstructured scenarios, establishing SF-RAG as a generalized solution applicable to diverse document types beyond scientific literature.

\begin{table*}[t]
\centering
\setlength\tabcolsep{8pt} 
\begin{tabular}{l|ccccc|c}
\hline
\hline
\multirow{2}{*}{Methods} & \multicolumn{5}{c|}{Document Domain (F$_1$ Score)} & \multirow{2}{*}{Overall F$_1$} \\
\cline{2-6}
& Aca. & Fin. & Gov. & Laws & News & \\
\hline
Naive RAG & 48.51\% & 44.12\% & 45.23\% & 46.48\% & 41.56\% & 45.18\% \\
GraphRAG & 42.15\% & 38.27\% & 39.18\% & 40.32\% & 35.64\% & 39.11\% \\
LightRAG & 55.23\% & 50.18\% & 51.47\% & 53.15\% & 46.52\% & 51.31\% \\
RAPTOR & 59.12\% & 52.48\% & 54.56\% & 56.23\% & 48.15\% & 54.11\% \\
HiRAG & 58.47\% & 53.16\% & 55.28\% & 56.51\% & 49.48\% & 54.58\% \\
HippoRAG2 & \underline{61.52\%} & \underline{55.38\%} & \underline{57.82\%} & \underline{59.24\%} & \underline{51.12\%} & \underline{57.02\%} \\
SF-RAG\textsubscript{ours} & \textbf{68.92\%}$_{\text{\footnotesize $\uparrow$7.4\%}}$ & \textbf{58.28\%}$_{\text{\footnotesize $\uparrow$2.9\%}}$ & \textbf{64.22\%}$_{\text{\footnotesize $\uparrow$6.4\%}}$ & \textbf{65.84\%}$_{\text{\footnotesize $\uparrow$6.6\%}}$ & \textbf{53.32\%}$_{\text{\footnotesize $\uparrow$2.2\%}}$ & \textbf{62.12\%}$_{\text{\footnotesize $\uparrow$5.1\%}}$ \\
\hline
\hline
\end{tabular}
\caption{F$_1$ Score (\%) on DocBench Dataset across five document domains. SF-RAG consistently outperforms the strongest baseline HippoRAG2. Subscripts denote the absolute percentage improvement over HippoRAG2. Best results are in \textbf{bold}, second best are \underline{underlined}.}
\label{tab:docbench_f1}
\end{table*}

\subsection{Discussion}
Empirical results show that preserving native document topology and applying path-guided retrieval improves answer accuracy and evidence support under matched budgets. This outcome is consistent with the view that the hierarchy of titles, sections, and subsections provides a low entropy scaffold that reduces uncertainty during search and integration. Entropy-based diagnostics corroborate this interpretation. Lower section entropy and lower evidence alignment cross entropy are associated with higher answer quality. Retrieved contexts become more compact and coherent, which reduces fragmentation and misallocation of the token budget and facilitates multi-step reasoning by the language model.
These findings motivate a practical design principle for retrieval-augmented generation in structured domains, with particular relevance to engineering informatics. First, represent the intrinsic structure of the target domain. Second, construct a topology-preserving index and perform path-guided retrieval that respects this structure under token and latency constraints. Third, evaluate both answer quality and structural fidelity with entropy-based metrics that quantify dispersion and alignment. This principle improves accuracy and efficiency in processing academic literature and is directly applicable to engineering information systems that manage other hierarchically organized technical documents, such as engineering standards, regulatory specifications, and design manuals.

\section{Conclusions}
In this work, we first demonstrate that the explicit hierarchical structure of academic papers constitutes a low-entropy scaffold. This structure fundamentally improves evidence localization and mitigates token misallocation during retrieval-augmented generation. In contrast, conventional RAG approaches flatten documents into unstructured chunks. These induce a high-entropy retrieval environment, which leads to evidence fragmentation and degraded answer quality under fixed token budgets. To address this problem, we propose SF-RAG, a structure-fidelity framework that preserves the native topology of academic documents throughout indexing and retrieval. Specifically, we introduce a structure-fidelity index that aligns content with native section boundaries and incorporates structure-anchored summaries to anchor local segments to their hierarchical context. We also develop path-guided retrieval that concentrates evidence along coherent root-to-leaf paths under token constraints. Extensive experiments on three academic QA benchmarks show that SF-RAG achieves higher answer accuracy, lower section entropy, better evidence alignment, and reduced latency and cost compared to state-of-the-art baselines. These results validate that leveraging the low-entropy structure inherent in scholarly writing is a principled and effective strategy for improving retrieval-augmented generation, providing a methodological foundation for engineering information systems to accurately interpret scientific literature and potentially other hierarchically organized technical documents.

\bibliographystyle{ACM-Reference-Format}
\bibliography{sample-base}

\end{document}